\documentstyle[12pt]{article}
\begin{document}

\centerline{{\em Submitted to Physical Review {\bf C}}}

\begin{flushright} {
PITT-RHI-95-007a\\
May, 1995
}
\end{flushright}

\vspace{2cm}
\begin{center}
{\bf
HBT ANALYSIS OF ANISOTROPIC TRANSVERSE FLOW
}\\
\bigskip
\bigskip
S. A. VOLOSHIN\footnote{ On leave from Moscow Engineering Physics Institute,
     Moscow, 115409,  Russia}
and W. E. CLELAND

\bigskip
{\em
University of Pittsburgh, Pittsburgh, PA 15260
}\\

\end{center}

\bigskip
{\footnotesize
\centerline{ABSTRACT}
\begin{quotation}
\vspace{-0.10in}
%

The effects of anisotropic transverse collective flow on
the HBT correlation function is studied.
There exist three different physics contributions
related to flow which affect the correlation function:
anisotropic source shape, anisotropic space-momentum
correlations in pion emission,
and the effects related to the HBT measurement of
the size of a moving source in different reference frames.
Resolution of these contributions experimentally can lead to a
detailed understanding of both collective flow in nucleus-nucleus collisions
and the  HBT technique itself.
A method is presented which permits the derivation of model
independent relations between the radius of a source measured
 in a frame in which it is moving and in its rest frame.

\vspace{7pt}
PACS number(s): 25.75.+r, 21.65.+f, 13.85.Ni
\end{quotation}}
\bigskip

\newpage
\section{Introduction}

The discovery of anisotropic transverse flow in nucleus
 nucleus collisions at
BNL AGS energies~\cite{l877prl} implies that some previous results
should be reevaluated taking into account the effects of flow.
One set of results in this category is the measurements
 of collision volumes
using the Hanbury-Brown--Twiss (HBT) technique.
In this paper we discuss how HBT results are  affected by flow.
In addition we show that the HBT study of nucleus nucleus collisions
can provide valuable information on the understanding of flow itself.

By anisotropic transverse flow we generally mean directed flow.
We restrict ourselves to the  consideration of symmetric collisions
of identical nuclei.
In this analysis we use the following geometrical definitions.
In the transverse plane we define as ``$in^+$'' the direction of
transverse flow; ``$in^-$'' is defined as the opposite direction;
and ``$perp$'' is the direction perpendicular to the ``$in$'' direction.
We use the coordinate system where the $x$ axis coincides
with ``$in^+$'' and the $z$ axis coincides with the beam direction.
The reaction plane is then the plane defined by the ``$in$'' direction
and the beam (the $x$-$z$ plane).
In this paper we consider only pion correlations (unless
stated explicitly otherwise), and we call a pion source simply a source.

We discuss the problem essentially on a qualitative level.
Our goal is to
find the physics which affects the HBT measurements, not to generate
a complete set of formulae to describe the general case.
In the discussion we keep in mind a heuristic picture of
pion production and try to understand how different
features of the production affect the HBT correlation function.
In this very simple picture pions are produced from
two sources different in origin.
``Direct'' pions are mostly produced in deep inelastic nucleon-nucleon
collisions in the zone where the nuclei overlap.
This source can be relatively small and is located close to
the center of the line joining the centers of the colliding nuclei.
The second source of pions in our picture is
 resonance (mostly delta) decays.
This source size should be close to the freeze-out nucleon radius.
In this naive picture it is clear that, for example, the ``geometry''
of the source of pions of different rapidities
(or different $p_t$, or different orientation of pion momenta
with respect to the reaction plane, etc.)
can be very different.
The effective source responsible for the emission of pions with a
given rapidity
can have nonzero longitudinal and transverse velocity.
Note that we use the picture only to illustrate the sensitivity of
the HBT function to different features of the source;
one should remember that the real collision picture is much more
complex.

In general, the difference in the effective sources for pions
in different rapidity ($p_t$, etc.) regions can be treated mathematically
as a correlation between the momentum and space position of pion emission.
(Following many others we consider pion emission semi-classically,
and do not discuss the space-time quantum mechanical uncertainty).
For example, if low $p_t$ pions are produced mostly in the
totally expanded stage of the collision, the extracted source
size could be larger  than that evaluated by using high $p_t$ pions,
emitted from the hot stage.
In Section~\ref{ssmc} we discuss how these correlations appear
in the expression for the correlation function.

\section{Geometrical shape of pion source}

We start with the question of whether it is possible to observe different
source sizes looking at the emitting object from different directions.
The answer is definitely ``yes'', for several reasons.
One of them is simply the anisotropic source shape;
due to transverse directed flow the effective source could be extended
in the reaction plane.
Directly produced pions are emitted mostly from the ``center''
(of the transverse plane), while pions from
$\Delta$ (which undergo the collective motion to a greater extent)
decays could have an ``off-center'' origin.
Thus the source size appears to be different if measured
from ``in'' or ``perp'' directions.

We use RQMD 1.08~\cite{lrqmd} generated events to evaluate the approximate
magnitude of the effect.
We study production point distributions of pions with rapidity
(in the laboratory frame) around
$y_{lab} \approx 3$ which is close to beam rapidity in Au+Au collisions
for a projectile energy of 11.4~$GeV$/nucleon.
We select pions which are emitted in the transverse plane in the
$+x$, $-x$, and $y$ directions, and analyze their distributions in
$x$, $y$, $z$, and $t$.
Pions are considered as being emitted in $+x$ direction if their
transverse momentum components satisfy the
condition $0<|p_y|/p_x<0.5$.
All calculations are performed in the center of mass (Au+Au) frame.
To suppress fluctuations related to very rare cases of decays of very
long lived resonances, we consider only pions produced within
the first 50~$fm/c$ after the collision.
The results are summarized in Table~1.

{}From the first moments, one can see that the center of the observed pion
source is shifted from the center line in the direction of flow.
The variances of the distributions can be considered as squares of the
effective source sizes in different directions.
These are the sizes which one would measure if the pions were to carry
the information on their production points (this is not the case
in reality!).
Due to the collision symmetry with respect to the reaction plane
some coefficients in Table~1
(such as $\langle y\rangle$ for the $+x$ and $-x$ cases) are
 expected to be zero
and others (such as $\langle x\rangle$ for $+y$ and $-y$ cases)
are expected to be equal.
The values actually obtained in the calculation can be used to estimate
roughly the uncertainty in the results.

The effective source velocities for each of the cases of
$+x$, $-x$, $+y$, and $-y$ directions of pion emission
can be estimated from Table~1 using the formula:
\begin{equation}
v_i = \frac{\langle r_i t \rangle
        -\langle r_i\rangle \langle t \rangle }
	{\langle  t^2\rangle-\langle t\rangle^2},
\end{equation}
which gives, for example,
$v^{\{+x\}}_x \approx 0.13$,
$v^{\{-x\}}_x \approx -0.08$,
$v^{\{+y\}}_y \approx 0.10$,
and for all direction of pion emission
$v_z \approx 0.76$.
Note the difference between the magnitudes of
$v^{\{+x\}}_x$ and  $v^{\{-x\}}_x$ which is just the difference
in transverse flow velocities due to anisotropic flow.
In collisions at this energy, flow is mostly carried by nucleons.
Pions are involved in flow mainly through baryon resonance decays.
It is important for our analysis that the pions from $\Delta$ decays
almost do not ``remember'' the flow,
since the change in pion momenta due to flow is small.
The physical reason is the same as that for the low $p_t$ enhancement.
When the resonance decays, the products share the resonance momentum
in proportion to the ratio of their masses.
The pion from $\Delta \rightarrow \pi p$ decay carries only about 1/7 of
the momentum of the $\Delta$.
Thus we can expect for pions only about an extra 10~$MeV/c$
of transverse momentum
in the direction of flow, if one takes reasonable values for
baryon flow~\cite{lab}.
Analysis of experimental data~\cite{lpbm} yields an average value of
$v_{t} \approx $0.3;
for the longitudinal component it gives $v_{long} \approx 0.5$.
These values could be different from our estimate due to the particular
region of pion phase space considered.

The HBT analysis in principle permits to investigate the source size in
many dimensions, studying the dependence of the correlation function
on different components of the relative pion pair momentum.
Usually the so called ``long-side-out'' coordinate system is used for
such an analysis.
We remind the reader that the ``long-side-out'' coordinate system is
defined in the following way (see Fig.~\ref{fsol}).
``Long'' is the direction of the beam;
``out'' is the direction of the pion pair transverse momentum; ``side'' is
the direction in the transverse plane perpendicular to the ``out''
direction.
In the case of anisotropic flow the picture becomes more
complicated; one must also take into account the direction of flow.
Consequently, we denote, for example, by ``$in^+$-out''
the ``out'' direction for the case when the pion pair momentum
points in the  direction of flow;
and ``$in^-$-side'' source size would mean the mean size of the source
in the ``side'' direction
as it appears for pions emitted in the direction opposite to flow.

If one considers something like ``$in$-out'' and ``$in$-side'' source sizes,
many interesting possibilities appear.
One can see from Table~1 that, for example,
the $+x$ and $-x$ effective sizes are quite different.
One of the reasons for this is shadowing.
Direct pions are produced most often in the region where
colliding nuclei overlap.
Let us look at the collision from the flow (``$in^+$'') direction,
where one expects the spectators and most
 of the nucleons of the projectile.
Certainly these nucleons distort the real image of the pion source.
{}From this  direction one can see mostly the pions from
the nucleon fireball and, consequently, one measures this fireball size.
For an observer from the opposite (in the transverse plane)
``$in^-$'' direction  the source image is not distorted by
shadowing so the picture could be quite different.
{}From this direction one sees simultaneously the hot
pion fireball and, spatially separated from it, the nucleon fireball.
The ``out'' size is the effective transverse size of the source
in the direction of the transverse momentum of the pion pair;
``$in^+$-out'' and ``$in^-$-out'' sizes should be
determined by the sizes of the two sources and their separation.
This could explain the difference in ``out'' sizes from Table~1
seen from ``$in^+$'' ($\sqrt{12.8}\,fm$) and ``$in^-$'' directions
($\sqrt{17.4}\,fm$).
``$In^+$-side'' and ``$in^-$-side'' sizes are the dimensions of the source
perpendicular to the direction of the pion momentum
as seen from ``$in^+$'' and ``$in^-$'' directions.
The difference in ``$in^+$-side'' and ``$in^-$-side'' sizes
($\sqrt{19.1}\,fm$ and $\sqrt{16.6}\,fm$ respectively) is mostly due
to the difference in sizes of nucleon and pion fireballs, not to their
separation.
Another reason for the difference could be the shadowing of the pion
source by nucleons in ``$in^+$'' case, which effectively results in
a larger observed source size.

If this picture is correct one could study very interesting
effects, varying the pseudorapidity of the correlated pair and
the relative (azimuthal) angle between the pion pair and  the reaction plane.
In connection with the arguments made above it is also interesting
to study in detail the pion triple differential
distributions calculated with respect to the reaction plane..
For example, the shadowing discussed above could result in
nonzero values of the third harmonic Fourier coefficient
of the pion azimuthal distribution.
The rapidity (pseudorapidity) dependence, if observed,
could provide information on the localization of dense nucleon matter.

It is very interesting to compare the results of pion interferometry
with the results of source size measurements by the two proton correlation
technique.
In our (oversimplified) picture the proton source
is different from the pion source.
The effects discussed in the next two sections are also different
for the two approaches, which makes the comparison more difficult but
also more interesting.

\section{HBT function and space-momentum correlations}\label{ssmc}

All of the arguments related to the difference
between the ``$in^+$'' and ``$in^-$'' sizes given above
can be formulated in terms of
the correlation between the pion momentum and the space-time location of
its production point.
The reason for a separate discussion (in the previous
section) is that the space-momentum correlations widely discussed
in the literature are generally related only to isotropic transverse flow
(any collective movement/phenomena
in multiparticle production we call flow in this paper).
The subject of our study is anisotropic flow.

Below we rederive the expressions for effective source radii
measured using the HBT technique, having in mind
that this derivation can be helpful in  understanding the flow
contribution to the correlation function.
Under a few simple assumptions the correlation function can be written
in the ``on-shell'' form~\cite{lpratt}:
\begin{equation}
C({\bf q},{\bf P})=1+
\frac {\int d^4x_1 \, d^4x_2\, S(x_1,{\bf P}/2) S(x_2,{\bf P}/2)
e^{-iq(x_1-x2)}}
{[\int d^4x\, S(x,{\bf P}/2)]^2},
\label{ecf}
\end{equation}
where $S(x,{\bf p})$ is the source function,
$q=p_1-p_2$ is the relative momentum, and
 $P=p_1+p_2$ is the total momentum
of the pair, $P=(E,{\bf P})$; $p_1$ and $p_2$ are pion 4-momenta.
By consideration of the exact form of the correlation function
 one can generate
the corrections~\cite{lchap} to the source parameters calculated
with Eq.~(\ref{ecf}).

The mathematics of the space-momentum correlation is very simple;
the correlation means that the source function does not factorize:
\begin{equation}
S(x,{\bf p}) \neq S_s(x)S_m({\bf p}).
\end{equation}
A consequences of this is that the values presented in Table~1
depend on the pion momentum.
We come to a trivial conclusion: the interferometry of
pions with definite momentum is sensitive to the source which
emits pions with just this momentum.
The measured source sizes are not the sizes of the whole source but only
of the effective region which emits the proper pions (the lengths of
homogeneity~\cite{lsin}).

To give an idea of how the correlation function depends on the
values discussed in the previous section and presented in Table~1,
we derive below the expressions for the source radii.
We consider the correlation function with a fixed value of ${\bf P}$.
In this case the correlation function depends only on three
variables.
{}From the fact that the pions are on the mass shell it follows that
\begin{equation}
q_0={\bf q} {\bf P} /E ={\bf q} {\bf V},
\end{equation}
where ${\bf V}$ is the velocity of the pion pair.
Then, defining $r=x_1-x_2=(t,{\bf r})$
the correlation function can be rewritten in the form:
\begin{equation}
C({\bf q},{\bf P})=1+
\frac{ \int d^4x_1 \, d^4x_2\, S(x_1,{\bf P}/2) S(x_2,{\bf P}/2)
e^{-i{\bf q}({\bf V} t -{\bf r})} }
{[\int d^4x\, S(x,{\bf P}/2)]^2}.
\end{equation}

The effective ``mean square source size'' can be defined~\cite{lbert}
through the second derivative of the correlation function with respect
to the component of ${\bf q}$ in the corresponding direction.
There are diagonal:
\begin{equation}
R_{i}^2=-1/2 (\partial^2C({\bf q})/\partial^2 q_i)|_{{\bf q} =0}
=\langle (V_i t-r_i)^2 \rangle /2,
\end{equation}
and cross terms~\cite{lchap}:
\begin{equation}
R_{ij}^2=-1/2 (\partial^2C({\bf q})/\partial q_i
\partial q_j )|_{{\bf q} =0}
=\langle (V_i t-r_i) (V_j t-r_j) \rangle/2 .
\end{equation}
The correlation function in this case reads:
\begin{equation}
C({\bf q} ,{\bf P} )=1 + \exp [-\sum_i q_i^2 R_i^2
-2\sum_{i \neq j}q_i q_j R^2_{ij}].
\end{equation}
The interpretation of the expressions for the radii is straightforward:
$R^2$ is the mean square of the
distance between pions at the moment the second pion being
produced (see also the discussion of this question in~\cite{lzaic}).

In the ``long-side-out'' coordinate
system
\begin{equation}
{\bf q} ({\bf V} t-{\bf r})=- q_{side}r_{side}
-q_{out}(V_{out}t-r_{out})-q_{long}(V_{long}t-r_{long}),
\end{equation}
and it follows that
\begin{equation}
2 R_{side}^2=\langle r_{side}^2 \rangle,
\end{equation}
\begin{equation}
2 R_{out}^2=\langle (r_{out}-V_{out}t)^2 \rangle=
\langle r_{out}^2 \rangle-2\langle r_{out}V_{out}t\rangle
+\langle (V_{out}t)^2 \rangle,
\end{equation}
\begin{equation}
2 R_{long}^2=\langle (r_{long}-V_{long}t)^2 \rangle=
\langle r_{long}^2 \rangle-2\langle r_{long}V_{long}t\rangle
+\langle (V_{long}t)^2 \rangle,
\end{equation}
\begin{equation}
2 R_{out,long}^2= \langle (r_{long}-V_{long}t)(r_{out}-V_{out}t) \rangle,
\end{equation}
where we have introduced the notation
\begin{equation}
\langle f \rangle =\frac{ \int d^4x_1 \, d^4x_2\, f\,
S(x_1,{\bf P}/2) S(x_2,{\bf P}/2) }
{[\int d^4x\, S(x,{\bf P}/2)]^2}.
\end{equation}
In the absence of flow
\begin{equation}
\langle r_{side} \rangle=0,
\end{equation}
due to azimuthal symmetry of the collision~\cite{lchap}.
For the same reason $R^2_{side,long}=R^2_{side,out}=0$.
For the case of non-zero flow these results are valid
only for ``in'' sizes due to the symmetry of
the collision with respect to  the reaction plane.
For the case of ``perp'' sizes all of the terms are in
the general case nonzero.

The expressions for the radii can be rewritten through the moments of one
particle  production point spatial distributions. For example:
\begin{eqnarray}
R_{x}^2&=&\langle (x-V_xt)^2 \rangle /2=
   \langle [(x_1-x_2)-V_x(t_1-t_2)]^2 \rangle /2 \nonumber \\
	&=&\langle (x_1-\langle x_1 \rangle) ^2\rangle
-2V_x (\langle x_1 t_1 \rangle -\langle x_1 \rangle \langle t_1 \rangle )
	+V_x^2 \langle (t_1-\langle t_1 \rangle) ^2\rangle,
\end{eqnarray}
and therefore the radii can be estimated using Table~1.
The results of radii calculation are presented in Table~2 taking into
account that in this case the mean transverse
and longitudinal velocities of pions
are approximately 0.38 and 0.87, respectively (in the center of mass of
colliding nuclei).

Due to the relatively small values of pion pair transverse
velocities the radii $R_{x}^2$ and $R_{y}^2$ are not very different from the
values of $\langle x^2\rangle-\langle x\rangle^2 $
and $\langle y^2\rangle-\langle y\rangle^2 $ in
Table~1. They still resemble the features of the true source geometry.
Due to the sizeable longitudinal pion velocity
the values of $R_{z}^2$ are very different from the values of
$\langle z^2\rangle-\langle z\rangle^2$.
Note also the possibility for non-zero values of cross-terms in the
radii matrix.
The relatively large value of $R_{xz}^2$ in the $-x$ case could indicate
spatial separation of the pion and nucleon fireballs.

\section{HBT measurements of a moving source}

The values given in Table~2 are the parameters of the HBT correlation
function as measured in the center-of-mass frame of the colliding nuclei.
As was shown in Section~2 the pion source in this frame has non-zero
collective longitudinal and transverse velocities.
Here we study how this motion distorts the HBT measurements
of the source sizes.

In some models it is possible to perform all necessary
calculations in covariant form by introducing the 4-velocity of the source.
However it is useful to derive model independent formulae by performing
Lorentz transformations between different systems.
Our goal is to establish the relationship between
the source size in its rest frame
and the HBT correlation function measured in the frame where
the source is moving.
The correlation function by definition is the ratio of the two-particle
invariant density and the product of the invariant one particle densities;
it is invariant under Lorentz transformations.
Using this property one can write:
\begin{equation}
C({\bf q},{\bf P})=C({\bf q}',{\bf P}')=
1+\frac{\int d^4x'_1 \, d^4x'_2\, S(x'_1,{\bf P}'/2) S(x'_2,{\bf P}'/2)
e^{-iq'(x'_1-x'_2)}}
      {[\int d^4x'\, S(x',{\bf P}'/2)]^2}
\label{ecf2}
\end{equation}
where the integrals are evaluated in the source rest frame,
and the prime denotes the coordinate and momentum values in this system.

Let us assume that our source moves with velocity ${\bf v}$.
We start with the case $v^2 \ll 1$. In this case the momentum
transformation equations are very simple:
\begin{equation}
q'_0=q_0-{\bf v} {\bf q} ={\bf q} ({\bf V} - {\bf v}),
\end{equation}
\begin{equation}
{\bf q}' = {\bf q} -{\bf v} q_0 = {\bf q} - {\bf v} ({\bf q} {\bf V}),
\end{equation}
\begin{equation}
{\bf P}'={\bf P} - {\bf v} E.
\end{equation}
Then
\begin{equation}
q'(x'_1-x'_2)=q' x'=({\bf q} ({\bf V} - {\bf v}))t' -
({\bf q} -{\bf v} ({\bf q} {\bf V})) {\bf r}' .
\end{equation}
One evaluates the second derivatives of the expression (\ref{ecf2})
with respect to $q_i$ to obtain the radii.
For example:
\begin{equation}
R^2_{x}=-1/2 (\partial^2C({\bf q})/\partial^2 q_x)|_{{\bf q} =0}=
\langle [t'(V_x-v_x) - x' + V_x ({\bf r}' {\bf v})]^2 \rangle '/2,
\end{equation}
\begin{equation}
R^2_{xy}=\langle [t'(V_x-v_x) - x' + V_x ({\bf r}' {\bf v})]
            [t'(V_y-v_y) - y' + V_y ({\bf r}' {\bf v})] \rangle '/2.
\end{equation}
The other radii can be computed in an analogous way.
The prime on the bracket ($\langle \rangle '$)
 means that the mean value is evaluated
using $S(x_1,{\bf P}'/2)$ instead of $S(x_1,{\bf P}/2)$.
In fact the source function depends on ${\bf P}$ rather slowly and  for
an estimate of the radii one can neglect the difference for the
case of small source velocity, which we consider here.
Note that  Eqs.~22 and 23 show explicitly the dependence of
the correlation function radii parameters on source velocity ${\bf v}$.

The physical interpretation of Eqs.~22 and 23 is the same as for the case
of the source being at rest, as can be shown by performing
an approximate Lorentz transformation.
The HBT correlation function measures the distance
between the pions at the moment of production of the second pion,
which is:
\begin{equation}
t{\bf V}-{\bf r} =
(t'+{\bf v}{\bf r}'){\bf V}-({\bf r}'+{\bf v}t')=
t'({\bf V}-{\bf v}) - {\bf r}' +  {\bf V}({\bf r}' {\bf v}),
\end{equation}
as calculated to the first order of $v$.

For the case of $v \ll 1$ the values of the second
moments of the spatial and temporal
distributions  are very close to each other in the both systems
(note that in the source rest frame the quantities
$\langle t'{\bf r}' \rangle  $ are zero).
Taking this into account and using the values from Table~1 one can estimate
the distortion of the correlation function due to transverse directed flow
by considering Lorentz boosts only in the transverse plane.
We find, for example, that about half of the difference between
$R_x^2$ in $+x$ and $-x$ cases is due to the transverse flow, with the
remaining part attributable to the difference in source geometry for
the two cases.

We consider the case of arbitrary $v$ and
use the exact Lorentz transformations for one particular case when
the flow velocity and pair velocity are directed along the $x$ axis.
In this case:
\begin{equation}
q'_0=\frac{q_0-vq_x}{\sqrt{1-v^2}}=\frac{q_xV-vq_x}{\sqrt{1-v^2}}=
q_x\frac{V-v}{\sqrt{1-v^2}},
\end{equation}
\begin{equation}
q'_x=\frac{q_x-vq_0}{\sqrt{1-v^2}}=\frac{q_x-vVq_x}{\sqrt{1-v^2}}=
q_x\frac{1-vV}{\sqrt{1-v^2}},
\end{equation}
where $V$ is the pion pair velocity, and $v$ is the flow velocity.
{}From this formula and expressions for $q_0'$ and $q_x'$
one can derive that:
\begin{equation}
R_{x}^2=\langle (\frac{t'(V-v)}{\sqrt{1-v^2}} -
       \frac{x'(1-vV)}{\sqrt{1-v^2}})^2  \rangle '
    =\frac{\langle (t'(V-v)-x'(1-vV))^2 \rangle '}{1-v^2}.
\label{err}
\end{equation}
For a simpler case of instantaneous freeze-out ($t=0$) and
of a Gaussian source the analogous formula was derived earlier
in~\cite{lsin2}.
Formula~(\ref{err}) is more general; it gives the dependence
of HBT radii on the source velocity independent of any model.
For anisotropic transverse flow the flow velocity depends on the
orientation of pion momenta with respect to the reaction plane.
This results in different apparent radii for the ``$in^+$'',
``$in^-$'', and ``$perp$''
cases even if the real source geometry is the same in all cases.

\section{Conclusion}

Anisotropic transverse flow produces anisotropy both in the pion source
geometry and in the space-momentum correlation.
It is also affects the interferometry size measurements because of nonzero
effective velocity of the source.
These three are the main phenomena responsible for the dependence of
the HBT correlation function on the orientation of pion pair momentum
with respect to the reaction plane.
A detailed study of the experimental data and model predictions is necessary
to disentangle all three effects,
but  such studies could be beneficial to our understanding
of flow in nucleus-nucleus collisions and of the HBT technique itself.

\section*{Acknowledgments}

The authors thank A.~Makhlin, D.~Mi\'{s}kowiec, and Yu.~Sinyukov
for fruitful discussions of the HBT technique,
and Y.~Zhang for help in providing the RQMD events.


%
%

\newpage
\section*{Tables}

\begin{table}[htbp]
\caption[]{\footnotesize
The first  and second moments and correlation coefficients
of spatial and temporal distribution
of charged pions  in Au+Au collisions as seen from different directions.
Pions are required to lie in the rapidity interval $2.7<y_{lab}<3.2$
and have transverse momentum $0.14\,GeV<p_t<0.25\,GeV$.
The impact parameter $3.0\,fm<b<6.0\,fm$.
Units for the first moments are $fm$ and for the second moments $fm^2$.
}
\label{trq2}

\medskip
\begin{center}
\begin{tabular}{|c|c|c|c|c||c|c|c|c|c|} \hline
      & $+x$ & $-x$ & $+y$ & $+y$ &
& $+x$ & $-x$ & $+y$ & $-y$ \\
\hline
$\langle x\rangle$                     & 4.9   & -0.5  & 1.6  & 1.7   &
 $\langle xy \rangle -\langle x\rangle\langle y\rangle$
& 0.2   & 0.0   & 0.1  & -0.7  \\
$\langle y\rangle$                     & -0.2  & -0.2  & 3.0  & -3.0  &
 $\langle xz \rangle -\langle x\rangle\langle z\rangle$
& 9.3   & 1.6   & 7.3  & 9.1   \\
$\langle z\rangle$                     & 11.2  & 9.2   & 9.9  & 9.3   &
 $\langle xt \rangle -\langle x\rangle\langle t\rangle$
& 12.2  & -7.1  & 5.3  & 9.1   \\
$\langle t\rangle$                     & 19.7  & 16.7  & 18.2 & 17.3  &
 $\langle yz \rangle -\langle y\rangle\langle z\rangle$
& -2.4  & -2.1  & 4.7  & -2.5  \\
$\langle x^2\rangle-\langle x\rangle^2        $& 12.8  & 17.4
& 17.6 & 15.8  &
 $\langle yt \rangle -\langle y\rangle\langle t\rangle$& -1.6
& -3.0  & 9.3  & -8.6  \\
$\langle y^2\rangle-\langle y\rangle^2        $& 19.1  & 16.6
& 16.9 & 17.1  &
 $\langle zt \rangle -\langle z\rangle\langle t\rangle$& 72.8
& 68.4  & 74.7 & 72.7  \\
$\langle z^2\rangle-\langle z\rangle^2        $& 69.4  & 61.1
& 68.7 & 66.2  & & & & & \\
$\langle t^2\rangle-\langle t\rangle^2        $& 92.3  & 92.2
& 98.0 & 99.1  & & & & & \\
\hline
\end{tabular}
\end{center}
\end{table}

\begin{table}[htbp]
\caption[]{\footnotesize
Matrix of source radii (in $fm^2$) found using the HBT correlation
function measured from the different directions with respect to flow
in the center of mass system of the colliding nuclei.
The input parameters are taken from Table~1.
}
\label{trq}

\medskip
\begin{center}
\begin{tabular}{|c|c|c|c|c|c|c|} \hline
     &$R_{x}^2$&$R_{y}^2$&$R_{z}^2$&$R_{xy}^2$&$R_{xz}^2$&$R_{yz}^2$\\
\hline
$x+$ &   16.9   & 19.1     &  12.6    &  0.8     & -1.5     & -1.0     \\
$x-$ &   25.3   & 16.6     &  11.8    &  -1.1    &  3.3     &  1.2     \\
$y+$ &   17.6   & 24.0     &  12.9    &  -1.9    &  2.7     &  0.6     \\
$y-$ &   15.8   & 24.8     &  14.7    &   2.8    &  1.2     & -0.2     \\
\hline
\end{tabular}
\end{center}
\end{table}

\clearpage
\newpage

\section*{Figure Caption}

\begin{enumerate}

\item    \label{fsol}
The definition of transverse momenta in the ``long-side-out''
coordinate system.
\end{enumerate}


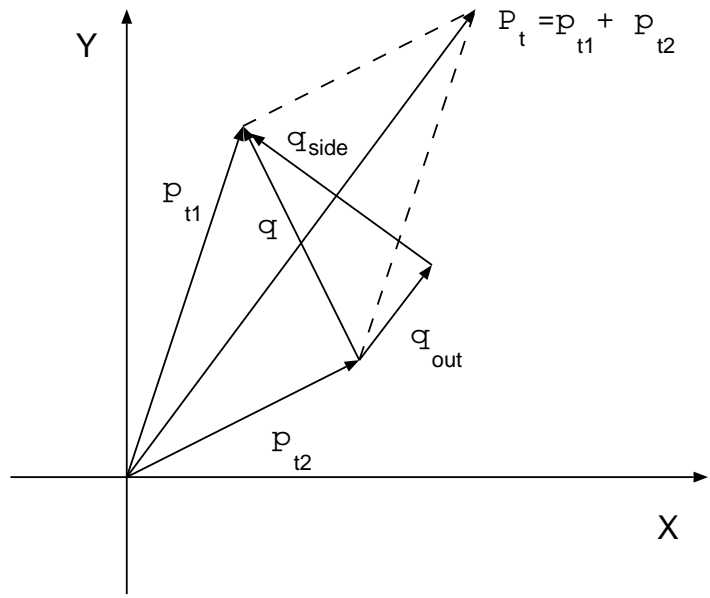
\end{document}